%% file: main.tex
\documentclass{llncs}
\usepackage{bbding}
\usepackage{graphicx}
\usepackage{afterpage}
\usepackage[caption=false]{subfig}
\usepackage{multirow}
\usepackage{multicol}
\usepackage{array}
\usepackage{xcolor}
\usepackage{float}
\usepackage[normalem]{ulem}
\usepackage{listings}
\usepackage{makecell}
\usepackage[hyphens]{url}
\usepackage{hyperref}
\usepackage{ulem}
\usepackage{tablefootnote}

\let\oldFootnote\footnote
\newcommand\nextToken\relax

\renewcommand\footnote[1]{%
    \oldFootnote{#1}\futurelet\nextToken\isFootnote}

\newcommand\isFootnote{%
    \ifx\footnote\nextToken\textsuperscript{,}\fi}
\newcolumntype{C}[1]{>{\centering\let\newline\\\arraybackslash\hspace{0pt}}m{#1}}
\newcolumntype{A}{>{\centering}m{0.588cm}}

\begin{document}
\pagestyle{headings} 

\title{Online Fault Classification in HPC Systems through Machine Learning}

 \author{
 Alessio Netti\inst{1,2}\Envelope, Zeynep Kiziltan\inst{1}, Ozalp Babaoglu\inst{1} \\ Alina S\^irbu\inst{3}, Andrea Bartolini\inst{4}, Andrea Borghesi\inst{4}}
        
 \institute{Department of Computer Science and Engineering, University of Bologna, Italy \\ 
     \email{\{alessio.netti, zeynep.kiziltan, ozalp.babaoglu\}@unibo.it}
     \and
     Leibniz Supercomputing Centre, Garching bei M\"unchen, Germany \\ 
     \email{alessio.netti@lrz.de}
     \and
     Department of Computer Science, University of Pisa, Italy \\ 
     \email{alina.sirbu@unipi.it}
     \and
     Department of Electrical, Electronic and Information Engineering\\ University of Bologna, Italy \\
     \email{\{a.bartolini, andrea.borghesi3\}@unibo.it}
}

\maketitle

\begin{abstract}
As \emph{High-Performance Computing} (HPC) systems strive towards the \emph{exascale} goal, studies suggest that they will experience excessive failure rates. For this reason, detecting and classifying faults in HPC systems as they occur and initiating corrective actions before they can transform into failures will be essential for continued operation. In this paper, we propose a fault classification method for HPC systems based on machine learning that has been designed specifically to operate with live streamed data. We cast the problem and its solution within realistic operating constraints of online use. Our results show that almost perfect classification accuracy can be reached for different fault types with low computational overhead and minimal delay. We have based our study on a local dataset, which we make publicly available, that was acquired by injecting faults to an in-house experimental HPC system.
\end{abstract}


\input{sections/Intro}
\input{sections/Workload}
\input{sections/Features}

\input{sections/Results}
\input{sections/Conclusions}

\bibliographystyle{splncs04}
\bibliography{main}

\end{document}

%% file: sections/Intro.tex
\section{Introduction}
\label{section:introduction}

\paragraph{Motivation.}
Modern scientific discovery is increasingly being driven by computation~\cite{villa2014scaling}. As such, HPC systems have become fundamental ``instruments'' for driving scientific discovery and industrial competitiveness. Exascale (\(10^{18}\) operations per second) is the moonshot for HPC systems and reaching this goal is bound to produce significant advances in science and technology. Future HPC systems will achieve exascale performance through a combination of faster processors and massive parallelism. With Moore's Law reaching its limit, the only viable path towards higher performance has to consider switching from increased transistor density towards increased core count, thus leading to increased failure rates~\cite{cappello2014toward}. Exascale HPC systems not only will have many more cores, they will also use advanced low-voltage technologies that are more prone to aging effects~\cite{bergman2008exascale} together with system-level performance and power modulation techniques, all of which tend to increase fault rates~\cite{engelmann2017resilience}. It is estimated that large parallel jobs will encounter a wide range of failures as frequently as once every 30 minutes on exascale platforms~\cite{snir2014addressing}. Consequently, exascale performance, although achieved nominally, cannot be sustained for the duration of applications running for long periods.

In the rest of the paper, we adopt the following terminology. A \emph{fault} is defined as an anomalous behavior at the hardware or software level that can lead to illegal system states (\emph{errors}) and, in the worst case, to service interruptions (\emph{failures})~\cite{gainaru2015errors}. Future exascale HPC systems must include automated mechanisms for masking faults, or recovering from them, so that computations can continue with minimal disruptions. This in turn requires detecting and classifying faults as soon as possible since they are the root causes of errors and failures.

\paragraph{Contributions.} 
We propose and evaluate a fault classification method based on supervised Machine Learning (ML) suitable for online deployment in HPC systems. Our approach relies on a collection of performance metrics that are readily available in most HPC systems. The experimental results show that our method can classify almost perfectly several types of faults, ranging from hardware malfunctions to software issues and bugs.
Furthermore, classification can be achieved with little computational overhead and with minimal delay, thus meeting real time requirements. 
We characterize the performance of our method in a realistic context similar to online use, where live streamed data is fed to fault classifiers both for training and for detection, dealing with issues such as class imbalance and ambiguous states. Most existing studies, on the contrary, rely on extensive manipulation of data, which is not feasible in online scenarios. Moreover, we reproduce the occurrence of faults basing on real failure traces.

Our evaluation is based on a dataset that we acquired from an experimental HPC system (called Antarex) where we injected faults using FINJ, a tool we previously developed~\cite{netti2018finj}. Making the Antarex dataset publicly available is another contribution of this paper. Acquiring our own dataset for this study was made necessary by the fact that commercial HPC system operators are very reluctant to share trace data containing information about faults in their systems~\cite{kondo2010failure}.

\input{sections/Relatedwork.tex}

\paragraph{Organization.} 
This paper is organized as follows. In Section~\ref{section:workload}, we describe the Antarex dataset, and in Section~\ref{section:features}, we discuss the features extracted from it. In Section~\ref{section:experimentalresults}, we present our experimental results, and we conclude in Section~\ref{section:conclusions}.

%% file: sections/Relatedwork.tex
\paragraph{Related Work.}
\label{section:relatedwork}
Automated fault detection through system performance metrics and fault injection has been the subject of numerous studies. However, ML-based methods using fine-grained monitored data (i.e., sampling once per second) are more recent. Tuncer et al.~\cite{tuncer2018online} propose a framework for the diagnosis of performance anomalies in HPC systems; however, they do not deal with faults that lead to errors and failures, which cause a disruption in the computation, but only with performance anomalies that result in longer runtimes for applications. Moreover, the data used to build the test dataset was not acquired continuously, but rather in small chunks related to single application runs. Thus it is not possible to determine the feasibility of this method when dealing with streamed, continuous data from an online HPC system. A similar work is proposed by Baseman et al.~\cite{baseman2016interpretable}, which focuses on identifying faults in HPC systems through temperature sensors. Ferreira et al.~\cite{ferreira2008characterizing} analyze the impact of CPU interference on HPC applications by using a kernel-level noise injection framework. Both works deal with specific fault types, and are therefore limited in scope.

Other authors have focused on using coarser-grained data (i.e., sampling once per minute) or on reducing the dimension of collected data, while retaining good detection accuracy. Bodik et al.~\cite{bodik2010fingerprinting} aggregate monitored data by using fingerprints, which are built from quantiles corresponding to different time epochs. Lan et al.~\cite{lan2010toward} discuss an outlier detection framework based on principal component analysis. Guan et al.~\cite{guan2012cda,guan2013adaptive} propose works focused on finding the correlations between performance metrics and fault types through a most relevant principal components method. Wang et al.~\cite{wang2010online} propose a similar entropy-based outlier detection framework suitable for use in online systems. These frameworks, which are very similar to threshold-based methods, are not suitable for detecting the complex relationships that may exist between different performance metrics under certain faults. One notable work in threshold-based fault detection is the one proposed by Cohen et al.~\cite{cohen2004correlating}, in which probabilistic models are used to estimate threshold values for performance metrics and detect outliers. This approach requires constant human intervention to tune thresholds, and lacks flexibility.

%% file: sections/Workload.tex
\section{The Antarex Dataset}
\label{section:workload}

The Antarex dataset contains trace data collected from an HPC system located at ETH Zurich while it was subjected to fault injections. The dataset is publicly available for use by the community and all the details regarding the test environment, as well as the employed applications and faults are extensively documented.\footnote{Antarex Dataset: \url{https://zenodo.org/record/2553224}} Due to space limitations, here we only give a short overview.

\subsection{Dataset Overview}
In order to acquire data, we executed benchmark applications and at the same time injected faults in a single compute node in the HPC system. The dataset is divided into two parts: the first includes only the CPU and memory-related benchmark applications and fault programs, while the second is strictly hard drive-related. We executed each part in both single-core and multi-core settings, resulting in 4 blocks of nearly 20GB and 32 days of data in total. The dataset's structure is summarized in Table~\ref{table:datasetstructure}. We acquired the data by continuous streaming, thus any study based on it will easily be reproducible on a real HPC system, in an online way.

\begin{table}[t!]
\centering
\caption{A summary of the structure for the Antarex dataset.}
\label{table:datasetstructure}
\fontsize{8.5}{8.5}\selectfont
\begin{tabular}{c|c|c|c|c|c}
\textbf{Dataset} & \textbf{Type}          & \textbf{Parallel} & \textbf{Duration} & \textbf{Benchmark} & \textbf{Fault} \\
 \textbf{Block} & & & & \textbf{Programs}  & \textbf{Programs} \\ \hline
Block I      & \multirow{2}{*}{CPU-Mem} & No       & \multirow{2}{*}{12 days}  & \multirow{2}{*}{\makecell{DGEMM, HPCC, \\ STREAM, HPL\tablefootnote{DGEMM: \url{https://lanl.gov/projects/crossroads/}, HPCC: \url{https://icl.cs.utk.edu/hpcc/}, STREAM: \url{https://www.cs.virginia.edu/stream/}, HPL: \url{https://software.intel.com/en-us/articles/intel-mkl-benchmarks-suite}}}}             & \multirow{2}{*}{\makecell{leak, memeater, ddot, \\ dial, cpufreq, pagefail}}        \\ \cline{1-1} \cline{3-3}
Block III    &  & Yes      &  & & \\ \hline
Block II     & \multirow{2}{*}{Hard Drive}    & No       & \multirow{2}{*}{4 days}   & \multirow{2}{*}{\makecell{IOZone, Bonnie++\tablefootnote{IOZone: \url{{https://iozone.org}}, Bonnie++: \url{https://coker.com.au/bonnie++/}}}}             & \multirow{2}{*}{\makecell{ioerr, copy}}         \\ \cline{1-1} \cline{3-3}
Block IV        & & Yes  & & &    \\       
\end{tabular}
\end{table}

\subsection{Experimental Setup for Data Acquisition}
The Antarex compute node used for data acquisition is equipped with two Intel Xeon E5-2630 v3 CPUs, 128GB of RAM, a Seagate ST1000NM0055-1V4 1TB hard drive and runs the CentOS 7.3 operating system. The node has a default Tier-1 computing system configuration.
The FINJ tool~\cite{netti2018finj} was used to execute benchmark applications and to inject faults, in a Python 3.4 environment. To collect performance metrics, we used the Lightweight Distributed Metric Service (LDMS) framework~\cite{agelastos2014lightweight}. We configured LDMS to sample a variety of metrics at each second, which come from the \emph{meminfo}, \emph{perfevent}, \emph{procinterrupts}, \emph{procdiskstats}, \emph{procsensors}, \emph{procstat} and \emph{vmstat} plugins. This configuration resulted in a total of 2094 metrics collected each second. Some of the metrics are node-level, and describe the status of the node as a whole, others instead are core-specific and describe the status of one of the 16 available CPU cores.

\subsection{Features of the Dataset}
\label{subsection:features}

 FINJ orchestrates the execution of benchmark applications and the injection of faults by means of 
 a workload file, which contains a list of benchmark and fault-triggering tasks to be executed at certain times, on certain cores, for certain durations. For this purpose, we used several FINJ-generated workload files, one for each block of the dataset. The details regarding the internal mechanisms driving FINJ are discussed in the associated work by Netti et al.~\cite{netti2018finj}.

\begin{figure*}[t!]
 \centering
 \captionsetup[subfigure]{}
  \subfloat[Histogram of fault durations.]{
    \includegraphics[width=0.48\textwidth,trim={0 5 10 5}, clip=true]{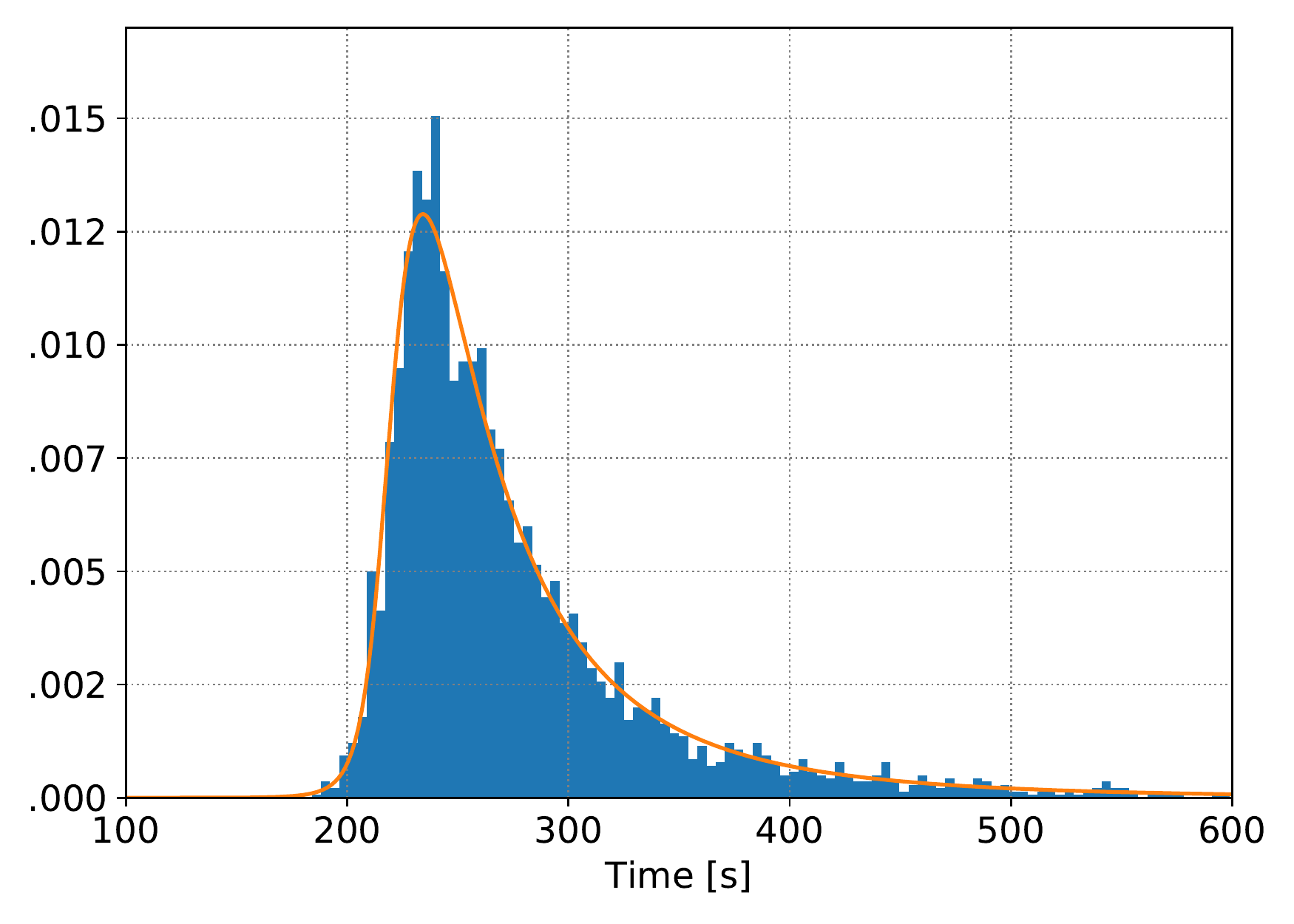}
  	}
  \subfloat[Histogram of fault inter-arrival times.]{
    \includegraphics[width=0.48\textwidth,trim={0 5 10 5}, clip=true]{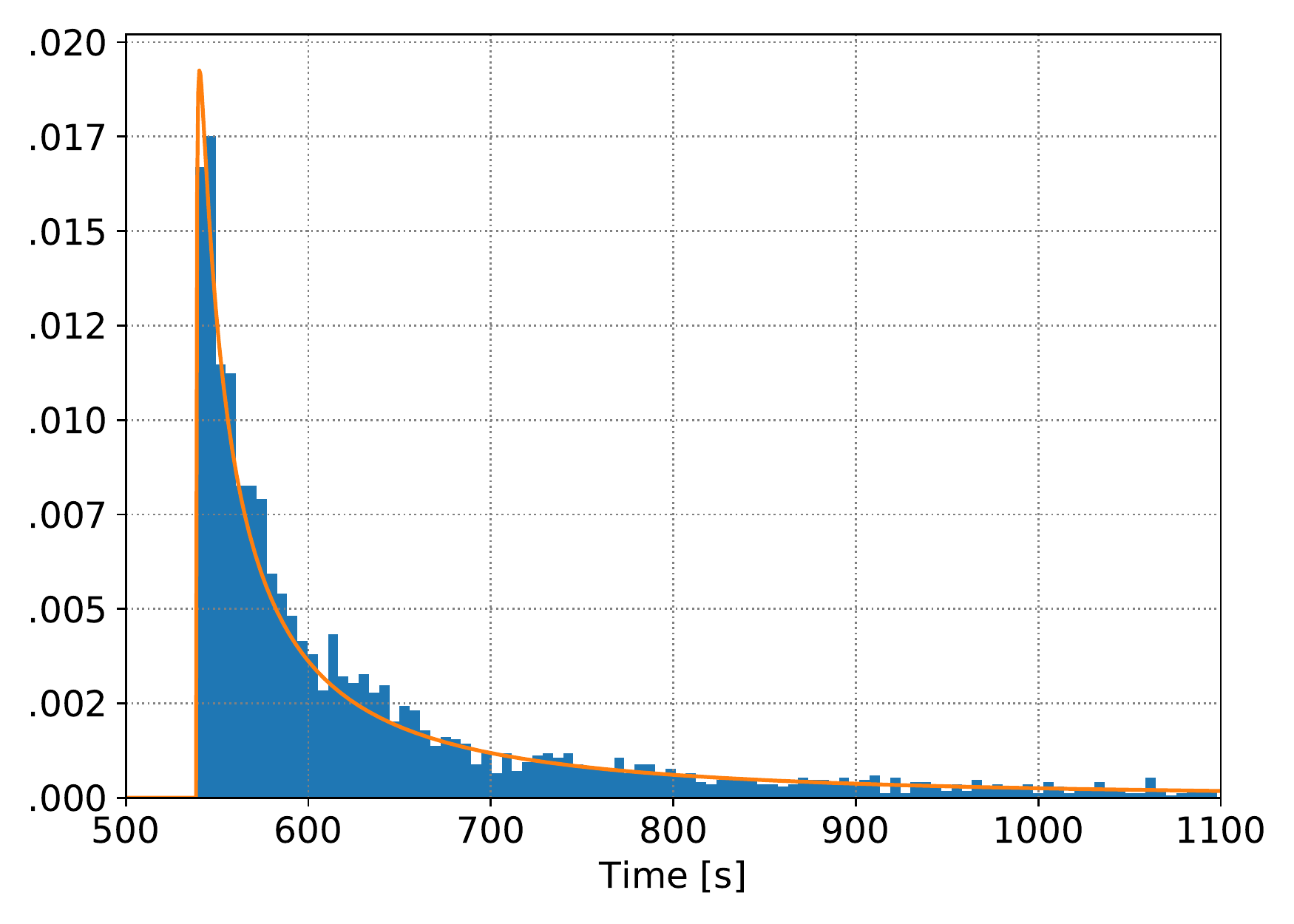}
  }
  \caption{Histograms for fault durations (a) and fault inter-arrival times (b) in the Antarex dataset compared to the PDFs of the Grid5000 data, as fitted on a Johnson SU and Exponentiated Weibull distribution respectively. We define the inter-arrival time as the interval between the start of two consecutive tasks.}
  \label{workload:pdf}
 \end{figure*}

\paragraph{Workload Files.} 
We used two statistical distributions in the FINJ workload generator to create the durations and inter-arrival times of the benchmark and fault-triggering tasks. The benchmark tasks are characterized by duration and inter-arrival times following normal distributions, and 75\% of the dataset's duration is spent running benchmarks. Fault-triggering tasks on the other hand are modeled using distributions fitted on the Grid5000 host failure trace available on the Failure Trace Archive.\footnote{Failure Trace Archive: \url{http://fta.scem.uws.edu.au/}} In Figure~\ref{workload:pdf}, we show the histograms for the durations (a) and inter-arrival times (b) of the fault tasks in the workload files, together with the original distributions fitted from the Grid5000 data.

FINJ generates each task in the workload by picking randomly the respective application to be executed, from those that are available. This implies that, statistically, all of the benchmark programs we selected will be subject to all of the available fault-triggering programs, given a sufficiently-long workload, with different execution overlaps depending on the starting times and durations of the specific tasks. Such a task distribution greatly mitigates overfitting issues. Finally, we do not allow fault-triggering program executions to overlap.

\paragraph{Benchmark Applications.} 
We used a series of well-known benchmark applications, stressing different parts of the node and providing a diverse environment for fault injection. Since we limit our analysis to a single machine, we use versions of the benchmarks that rely on shared-memory parallelism, for example through the OpenMP library. The benchmark applications are listed in Table~\ref{table:datasetstructure}.

\paragraph{Fault Programs.}
All the fault programs used to reproduce anomalous conditions on Antarex are available at the FINJ Github repository~\cite{netti2018finj}. As in~\cite{tuncer2018online}, each program can also operate in a low-intensity mode, thus doubling the number of possible fault conditions. While we do not physically damage hardware, we closely reproduce several reversible hardware issues, such as I/O and memory allocation errors. Some of the fault programs (\emph{ddot} and \emph{dial}) only affect the performance of the CPU core they run on, whereas the other faults affect the entire compute node. The programs and the generated faults are as follows.

\begin{enumerate}
\item \emph{leak} periodically allocates 16MB arrays that are never released~\cite{tuncer2018online} creating a \emph{memory leak}, causing memory fragmentation and severe system slowdown;
\item \emph{memeater} allocates, writes into and expands a 36MB array~\cite{tuncer2018online}, decreasing performance through a \emph{memory interference} fault and saturating bandwidth;
\item \emph{ddot} repeatedly calculates the dot product between two equal-size matrices. The sizes of the matrices change periodically between 0.9, 5 and 10 times the CPU cache's size~\cite{tuncer2018online}. This produces a \emph{CPU and cache interference} fault, resulting in degraded performance of the affected CPU;
\item \emph{dial} repeatedly performs floating-point operations over random numbers~\cite{tuncer2018online}, producing an \emph{ALU interference} fault, resulting in degraded performance for applications running on the same core as the program;
\item \emph{cpufreq} decreases the maximum allowed CPU frequency by 50\% through the Linux Intel P-State driver.\footnote{Intel P-State Driver: \url{https://kernel.org/doc/Documentation/cpu-freq}} This simulates a \emph{system misconfiguration} or \emph{failing CPU} fault, resulting in degraded performance;
\item \emph{pagefail} makes any page allocation request fail with 50\% probability.\footnote{Linux Fault Injection: \url{https://kernel.org/doc/Documentation/fault-injection}} This simulates a \emph{system misconfiguration} or \emph{failing memory} fault, causing performance degradation and stalling of running applications;
\item \emph{ioerr} fails one out of 500 hard-drive I/O operations with 20\% probability, simulating a \emph{failing hard drive} fault, and causing degraded performance for I/O-bound applications, as well as potential errors;
\item \emph{copy} repeatedly writes and then reads back a 400MB file from a hard drive. After such a cycle, the program sleeps for 2 seconds~\cite{guan2013adaptive}. This simulates an \emph{I/O interference} or \emph{failing hard drive} fault by saturating I/O bandwidth, and results in degraded performance for I/O-bound applications.
\end{enumerate} 

%% file: sections/Features.tex
\section{Creation of Features}
\label{section:features}

In this section, we explain how a set of features describing the state of the system for classification purposes was obtained from the metrics collected by LDMS.

\paragraph{Post-Processing of Data.} 
Firstly, we removed all constant metrics (e.g., the amount of total memory in the node), which were redundant, and we replaced the raw monotonic counters captured by the \emph{perfevent} and \emph{procinterrupts} plug-ins with their first-order derivatives. Furthermore, we created an \emph{allocated} metric, both at the CPU core and node level, and integrated it in the original set. This metric has a binary value, and defines whether there is a benchmark allocated on the node or not. Using such a metric is reasonable, since in any HPC system there is always knowledge of which jobs have computational resources currently allocated to them. Lastly, for each metric above, at each time point, we added its first-order derivative to the dataset as proposed by Guan et al.~\cite{guan2012cda}.

Feature vectors were then created by aggregating the post-processed LDMS metrics. Each feature vector corresponds to a 60-second aggregation window and is related to a specific CPU core. The step between feature vectors is of 10 seconds. This allows for high granularity and quick response times to faults. For each metric, we computed several indicators of the distribution of the values measured within the aggregation window~\cite{tuncer2018online}. These are the \emph{average}, \emph{standard deviation}, \emph{median}, \emph{minimum}, \emph{maximum}, \emph{skewness}, \emph{kurtosis}, and finally the \emph{5th, 25th, 75th} and \emph{95th percentiles}. This results in a total of 22 statistical features, including also those related to the first-order derivatives, for each metric in the dataset. The final feature vectors contain thus a total of 3168 elements. This number does not include the metrics collected by the \emph{procinterrupts} plugin, which were found to be irrelevant after preliminary testing. All the scripts used to process the data are available on the FINJ Github repository~\cite{netti2018finj}.

\paragraph{Labeling.} 
In order to train classifiers to distinguish between faulty and normal states, we labeled the feature vectors either according to the fault program (i.e., one of the 8 programs presented in Section \ref{subsection:features}) running within the corresponding aggregation window, or ``healthy'' if no fault was running. The logs produced by the FINJ tool, which are included in the Antarex dataset, detail the fault programs running at each time-stamp. In a generic deployment scenario, if users wish to perform training using data from spontaneous faults in the system, they need to provide the labels explicitly instead of relying on fault injection.

A single aggregation window may capture multiple system states, making labeling not trivial. For example, a feature vector may contain ``healthy'' time points that are before and after the start of a fault, or even include two different fault types. We define these feature vectors as \emph{ambiguous}. By using a short aggregation window of 60 seconds, we aim to minimize the impact of such ambiguous system states on fault detection. Since these cannot be completely removed, we experiment with two labelling methods. The first method is \emph{mode}, where all the labels that appear in the time window are considered. Their distribution is examined and the label appearing the most is used for the feature vector. This leads to robust feature vectors, whose label is always representative of the aggregated data. The second method is \emph{recent}, in which the label is given by the state of the system at the most recent time point in the time window. This could correspond to a fault type or could be ``healthy''. Such an approach may lead to a more responsive fault detection system, where what is detected is the system state at the moment, rather than the state over the last 60 seconds.

\paragraph{Detection System Architecture.}

\begin{figure}[t]
 \centering
  \includegraphics[width=0.9\textwidth,trim={0 0 0 0}, clip=true]{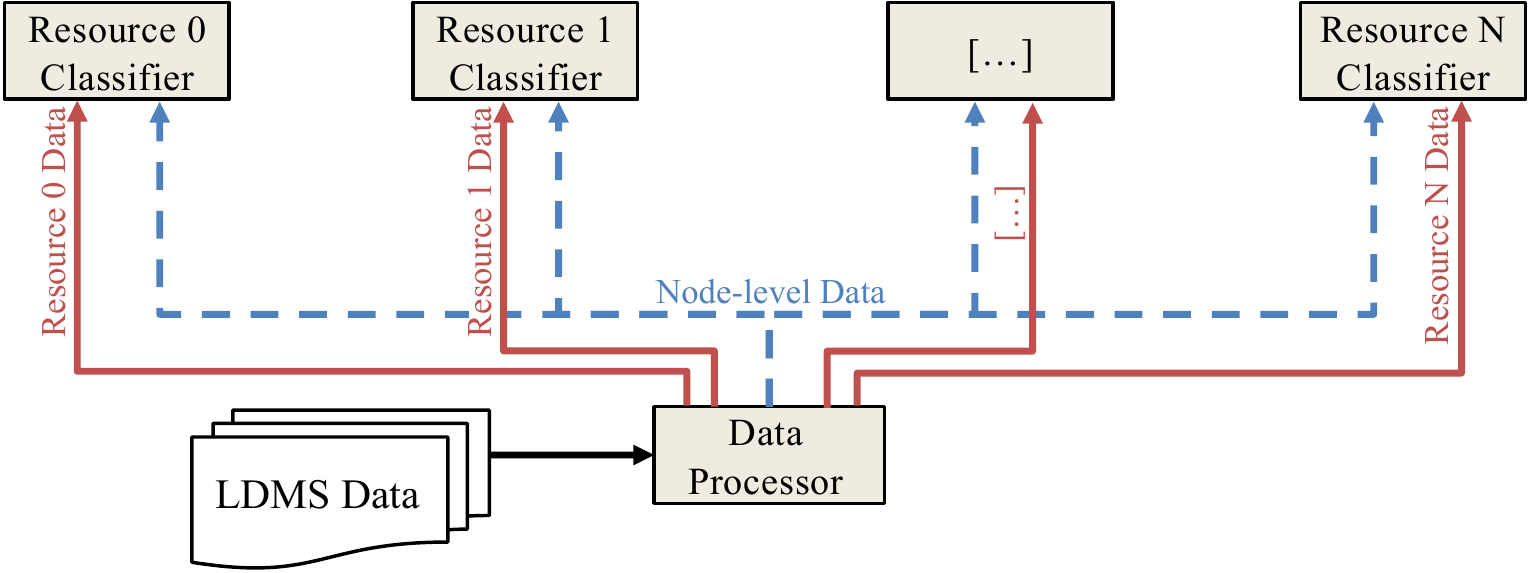}
  \caption{Architecture of our machine learning-based fault detection system.}
  \label{features:architecture}
 \end{figure}

For our fault detection system, we adopted an architecture based on an array of classifiers (Figure~\ref{features:architecture}). Each classifier corresponds to a specific computing resource type in the node, such as CPU cores, GPUs, MICs, etc. Each classifier is then trained with feature vectors related to all resource units of that type, and is able to perform fault diagnoses for all of them, thus detecting faults both at node level and resource level (e.g., dial and ddot). To achieve this, the feature vectors for each classifier contain all \emph{node-level} metrics for the system, together with \emph{resource-specific} metrics for the resource unit being considered. Since each feature vector contains data from one resource unit at most, this approach has the benefit of limiting the size of feature vectors, which improves overhead and detection accuracy. This architecture relies on the assumption that resource units of the same type behave in the same way, and that the respective feature vectors can be combined in a coherent set. However, users can also opt to use separate classifiers for each resource unit of the same type, overcoming this limitation, without any alteration to the feature vectors themselves. In our case, the compute node only contains CPU cores. Therefore, we train one classifier with feature vectors that contain both node-level and core-level data, for one core at a time.

The classifiers' training can be performed offline, using labeled data resulting from normal system operation or from fault injection (as in our case). The trained classifiers can then be deployed to detect faults on new, streamed data. Due to this classifier-based architecture, we can only detect one fault at any time. This design assumption is reasonable for us, as the purpose of our study is to distinguish between different fault scenarios automatically. In a real HPC system, although as a rare occurrence, multiple faults may be affecting the same compute node at the same time. In this case, our detection system would only detect the fault whose effects on the system are deemed more relevant by the classifier.

%% file: sections/Results.tex
\section{Experimental Results}
\label{section:experimentalresults}

We tested a variety of classifiers, trying to correctly detect which of the 8 faults described in Section \ref{subsection:features} were injected in the HPC node at any point in time of the Antarex dataset. The environment we used was Python 3.4, with the Scikit-learn package. We built the test dataset by picking the feature vector of only one randomly-selected core for each time point. Classifiers were thus trained with data from all cores, and can compute fault diagnoses for any of them.

We chose 5-fold cross-validation for evaluation of classifiers, using the average F-score as metric, which corresponds to the harmonic mean between precision and recall. When not specified, feature vectors are read in time-stamp order. In fact, while shuffling is widely used in machine learning as it can improve the quality of training data, such a technique is not well suited to our fault detection framework. Our design is tailored for online systems, where classifiers are trained using only continuous, streamed, and potentially unbalanced data as it is acquired, while ensuring robustness in training so as to detect faults in the near future. Hence, it is very important to assess the detection accuracy without data shuffling. We reproduce this realistic, online scenario by performing cross-validation on the Antarex dataset using feature vectors in time-stamp order. Most importantly, time-stamp ordering results in cross-validation folds, each containing data from a specific time frame. Only a small subset of the tests is performed using shuffling for comparative purposes. 

\subsection{Comparison of Classifiers}
\label{section:classifiers}

\begin{figure*}[t!]
 \centering
 \captionsetup[subfigure]{}
  \subfloat[Random Forest.]{
    \includegraphics[width=0.487\textwidth,trim={0 5 10 5}, clip=true]{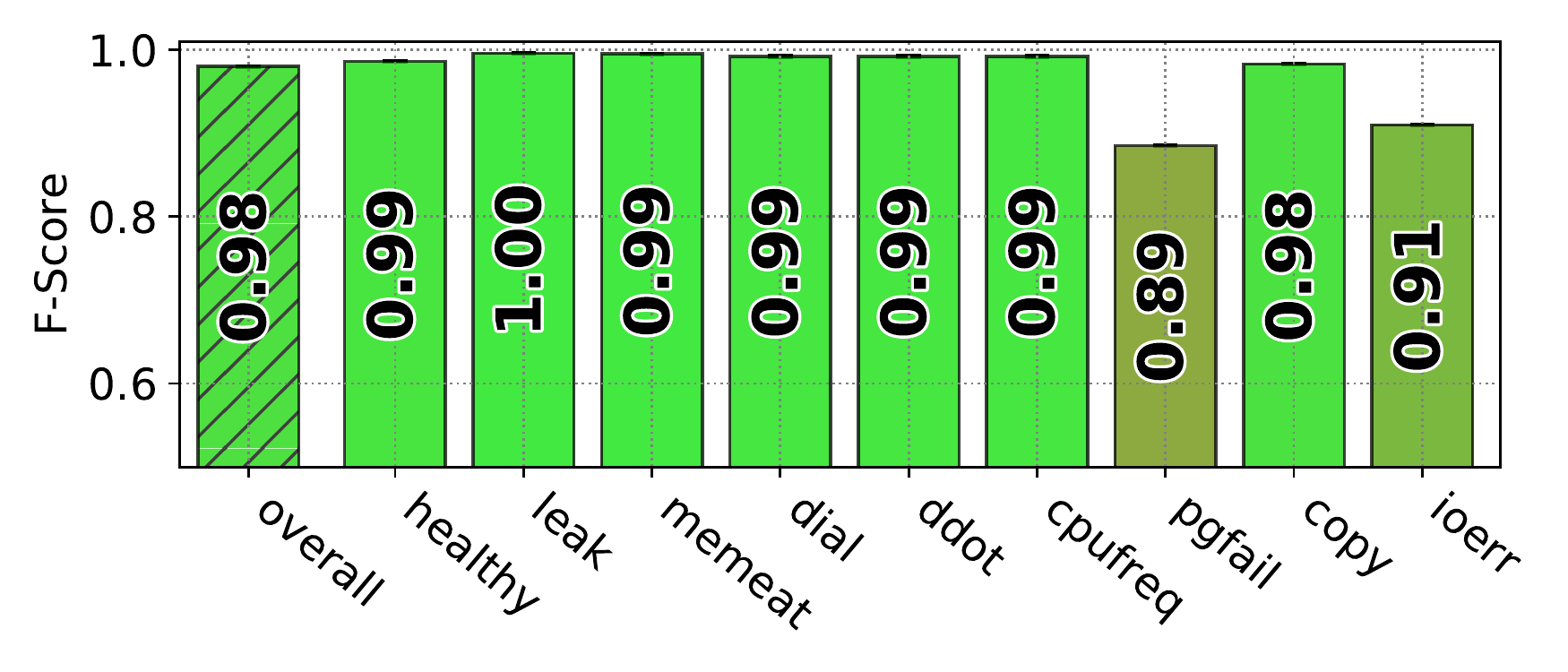}
  	}
  \subfloat[Decision Tree.]{
    \includegraphics[width=0.474\textwidth,trim={22 5 10 5}, clip=true]{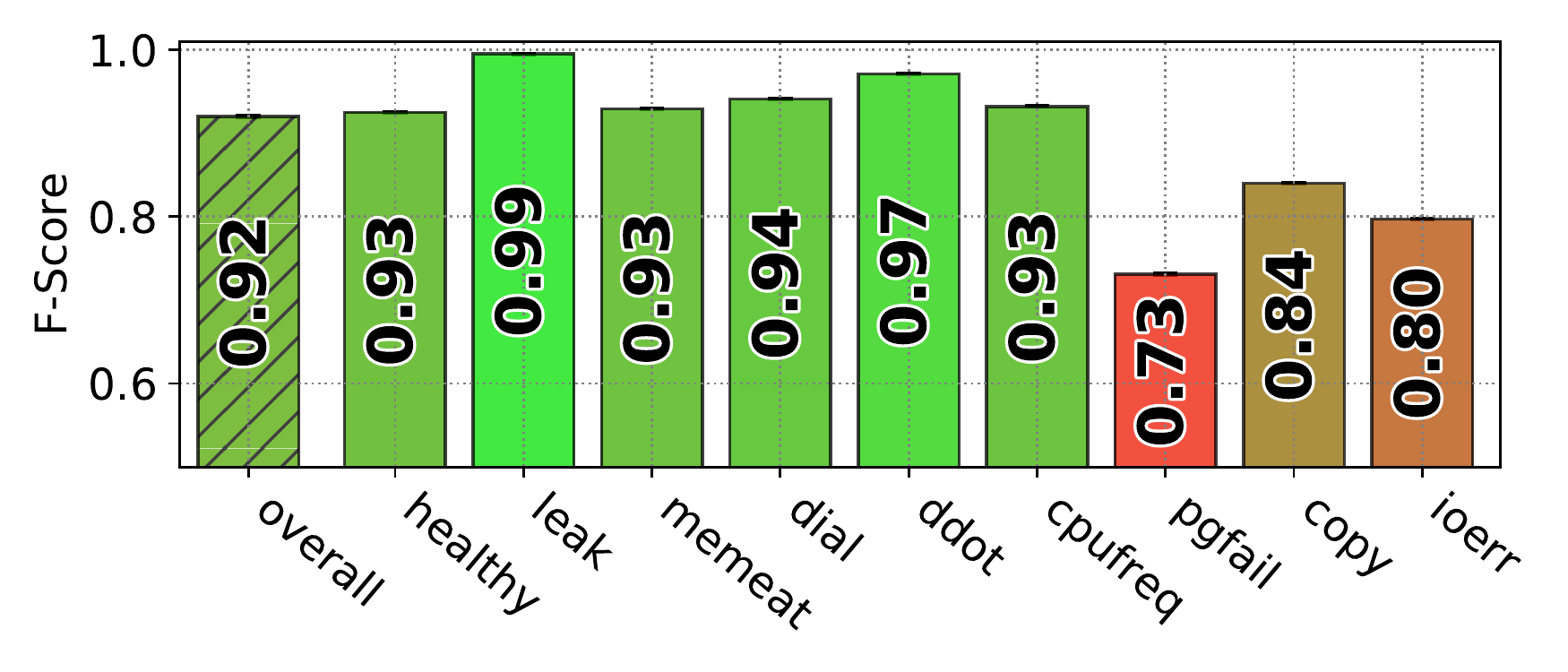}
  }
  \\
    \subfloat[Neural Network.]{
    \includegraphics[width=0.487\textwidth,trim={0 5 10 5}, clip=true]{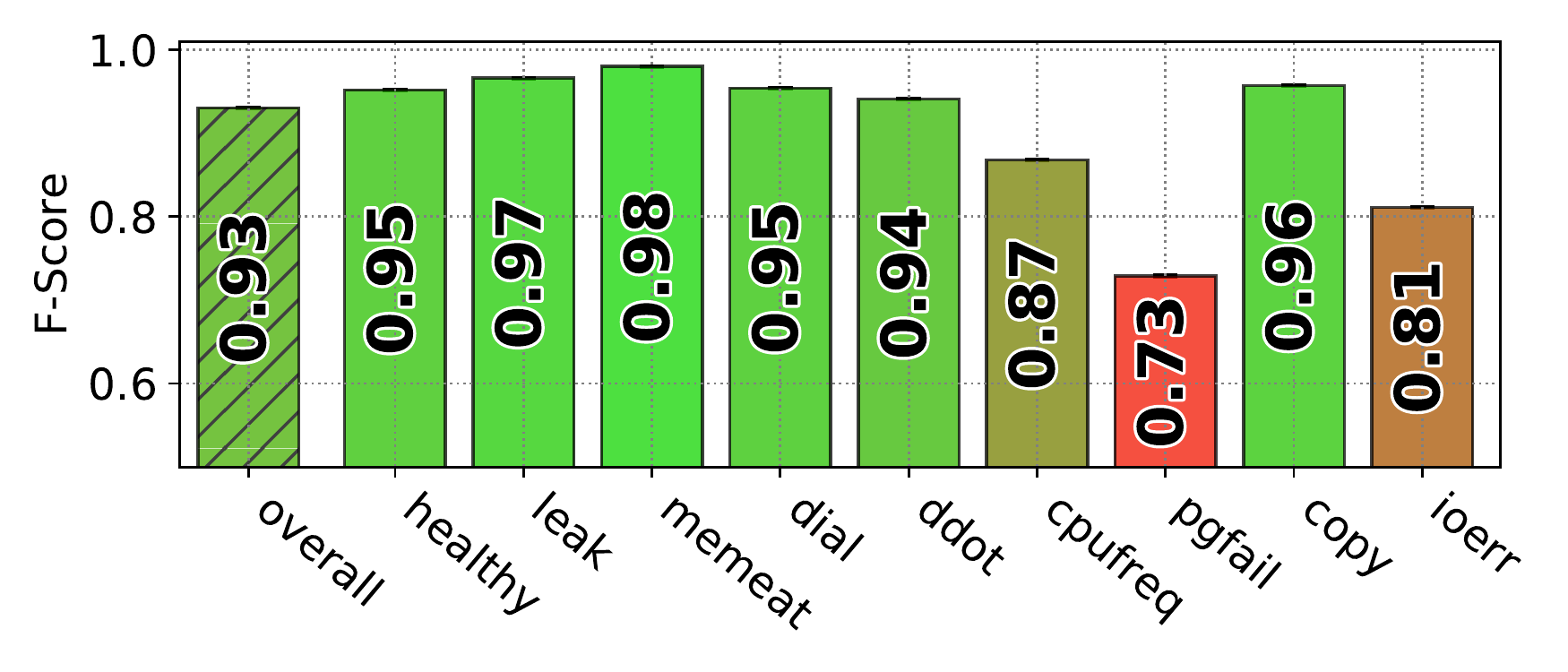}
  	}
  \subfloat[Support Vector Classifier.]{
    \includegraphics[width=0.474\textwidth,trim={22 5 10 5}, clip=true]{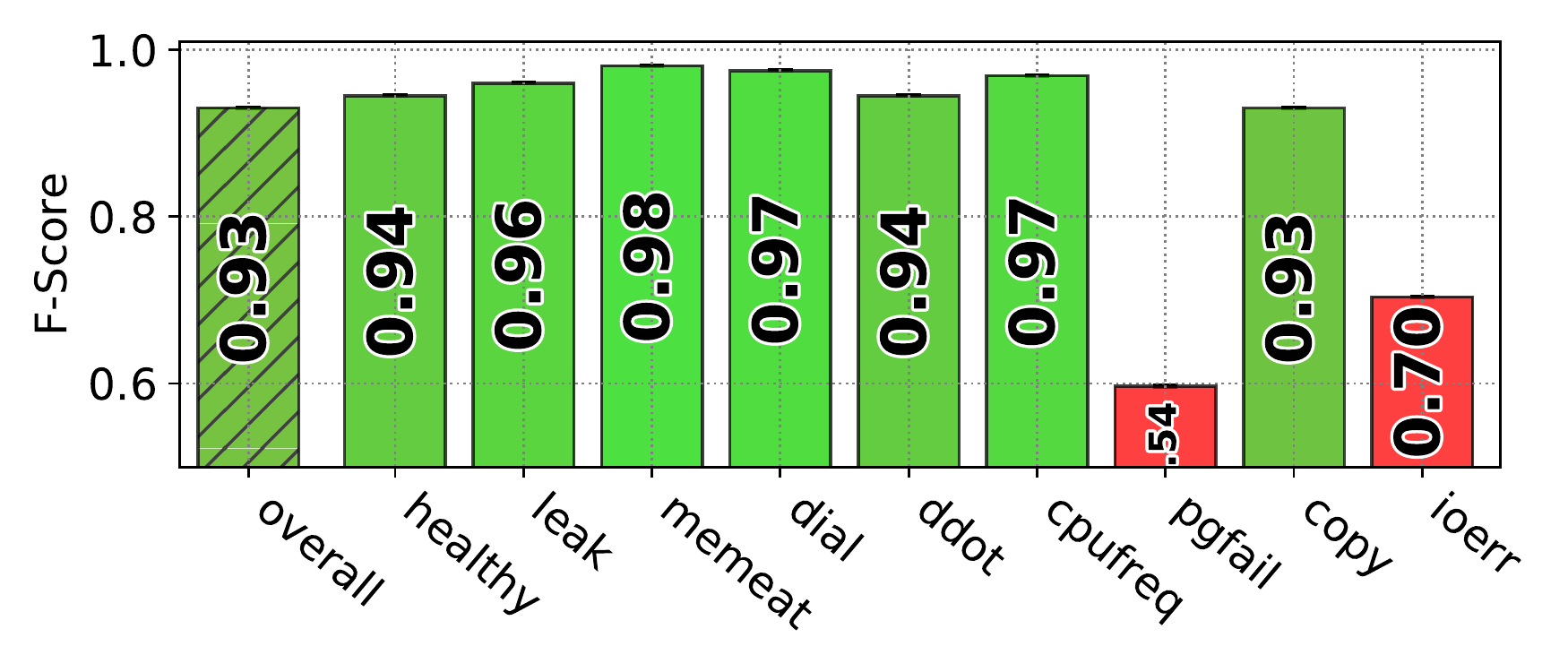}
  }
  \caption{The classification results on the Antarex dataset, using all feature vectors in time-stamp order, the \emph{mode} labeling method, and different classifiers.}
  \label{results:classifiers}
 \end{figure*}

For this experiment, we preserved the time-stamp order of the feature vectors and used the \emph{mode} labeling method. We included in the comparison a Random Forest (RF), Decision Tree (DT), Linear Support Vector Classifier (SVC) and Neural Network (MLP) with two hidden layers, each having 1000 neurons. We choose these four classifiers because they characterize the performance of our method well, and omit results on others for space reasons. The results for each classifier and for each class are presented in Figure~\ref{results:classifiers}. In addition, the overall F-score is highlighted for each classifier. It can be seen that all classifiers show very good performance, with F-scores that are well above 0.9. RF is the best classifier, with an overall F-score of 0.98, followed by MLP and SVC scoring 0.93. The critical point for all classifiers is represented by the \emph{pagefail} and \emph{ioerr} faults, which have substantially worse scores than the others. 

We infer that a RF would be the ideal classifier for an online fault detection system, due to its 5\% better detection accuracy, in terms of F-score, over the others. Additionally, random forests are computationally efficient, and therefore would be suitable for use in online environments with strict overhead requirements. It should be noted that unlike the MLP and SVC classifiers, RF and DT did not require data normalization. Normalization in an online environment is hard to achieve, as many metrics do not have well-defined upper bounds. To address this issue, a rolling window-based dynamic normalization approach can be used~\cite{guan2013adaptive}. This approach is unfeasible for ML classification, as it can lead to quickly-degrading detection accuracy and to the necessity of frequent training. Hence, in the following experiments we will use a RF classifier. 

\subsection{Comparison of Labeling Methods and Impact of Shuffling}
 
\begin{figure*}[t!]
 \centering
 \captionsetup[subfigure]{}
    \subfloat[Mode labeling.]{
    \includegraphics[width=0.487\textwidth,trim={0 5 10 5}, clip=true]{figures/results_RF_NOSHUFFLE.pdf}
  	}
  \subfloat[Recent labeling.]{
    \includegraphics[width=0.474\textwidth,trim={22 5 10 5}, clip=true]{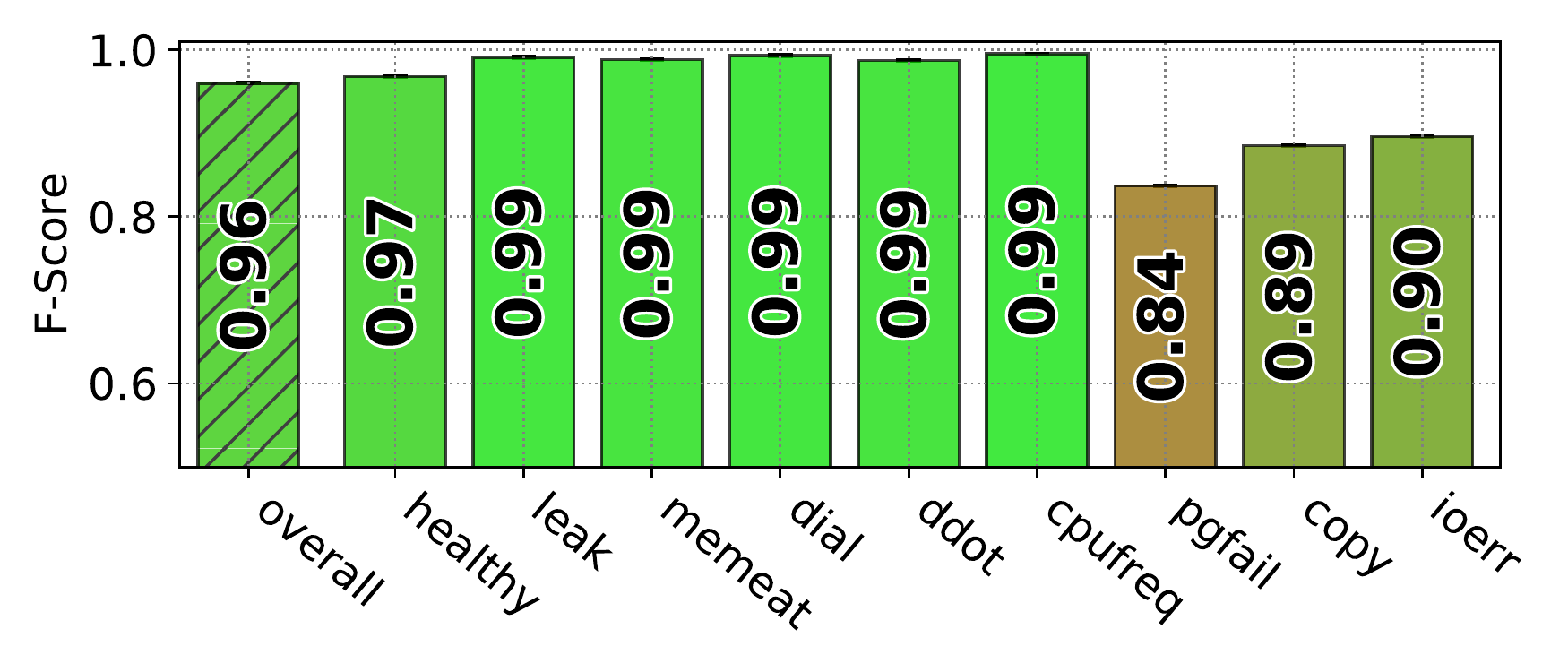}
  }
  \\
    \subfloat[Mode labeling with shuffling.]{
    \includegraphics[width=0.487\textwidth,trim={0 5 10 5}, clip=true]{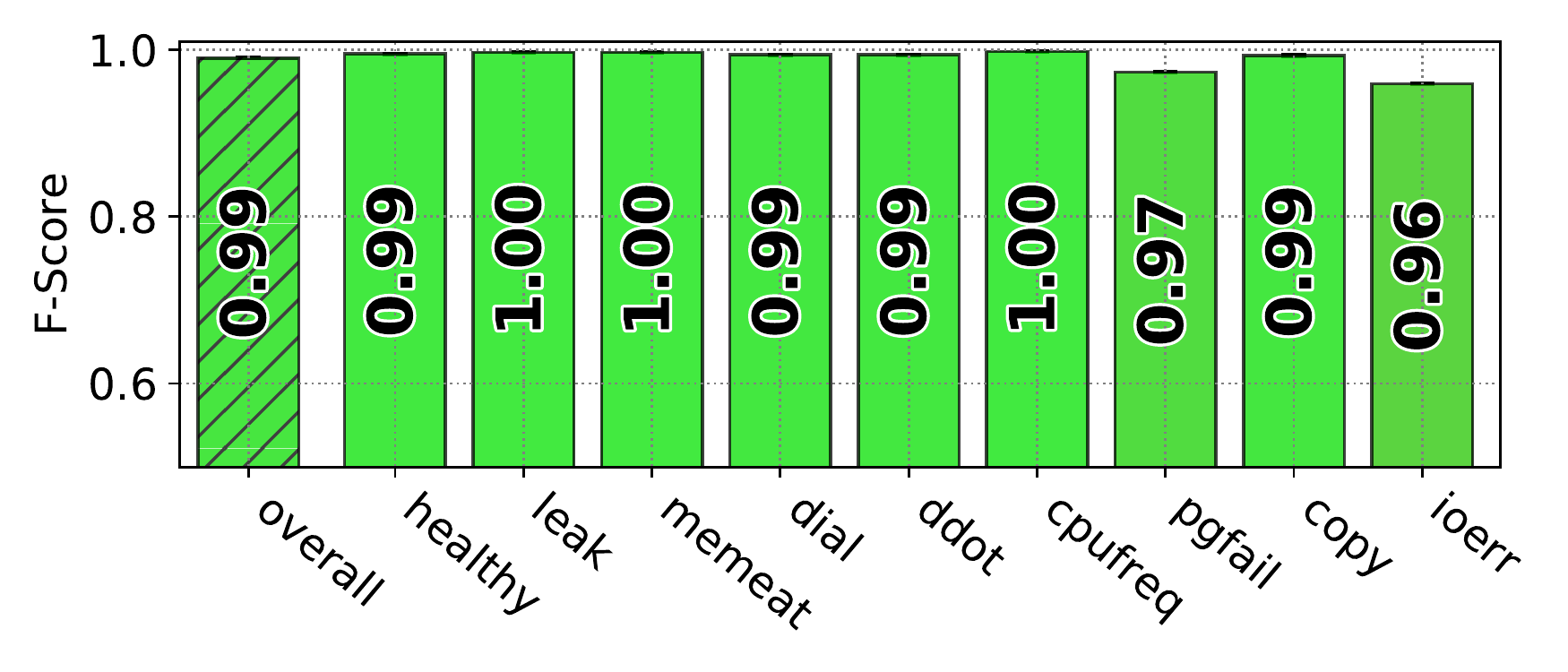}
  	}
  \subfloat[Recent labeling with shuffling.]{
    \includegraphics[width=0.474\textwidth,trim={22 5 10 5}, clip=true]{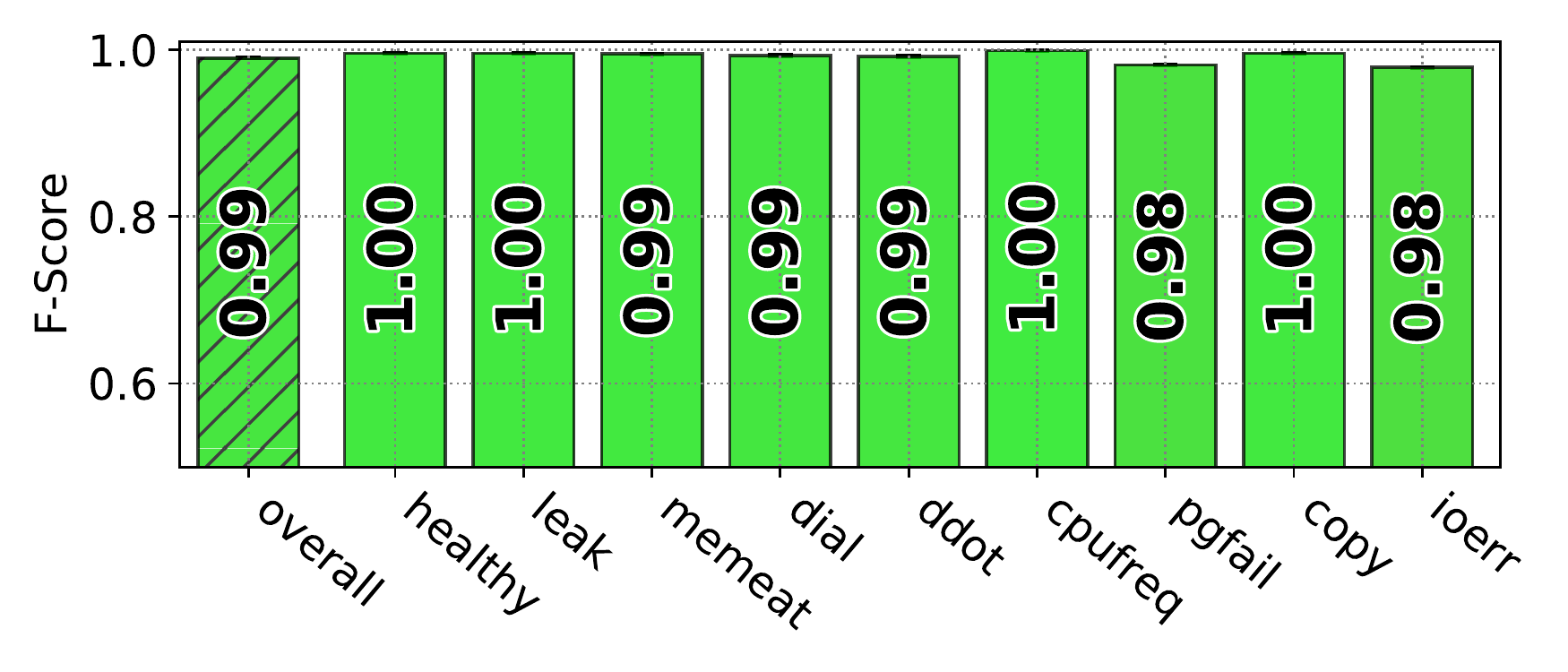}
  }
  \caption{RF classification results, using all feature vectors in time-stamp (top) or shuffled (bottom) order, with the \emph{mode} (left) and \emph{recent} (right) labeling methods.}
  \label{results:labeling}
 \end{figure*}

Here we evaluate the two different labeling methods we implemented by using a RF classifier. The results for classification without data shuffling can be seen in Figures~\ref{results:labeling}a for \emph{mode} and~\ref{results:labeling}b for \emph{recent}, with overall F-scores of 0.98 and 0.96 respectively, being close to the ideal values. Once again, in both cases the \emph{ioerr} and \emph{pagefail} faults perform substantially worse than the others. This is likely because both faults have an intermittent nature, with their effects depending on the hard drive I/O (ioerr) and memory allocation (pagefail) patterns of the underlying applications, proving more difficult to detect than the other faults.

In Figures~\ref{results:labeling}c and~\ref{results:labeling}d, the results with data shuffling enabled are presented for the \emph{mode} and \emph{recent} methods, respectively. Adding data shuffling produces a sensible improvement in detection accuracy for both of the labeling methods, which show almost ideal performance for all fault programs, and overall F-scores of 0.99. Similar results were observed with the other classifiers presented in Section~\ref{section:classifiers}, not shown here for space reasons. It can also be seen that in this scenario, \emph{recent} labeling performs slightly better for some fault types. This is likely due to the highly reactive nature of such labeling method, which can capture system status changes more quickly than the \emph{mode} method. The greater accuracy (higher F-score) improvement obtained with data shuffling and \emph{recent} labeling, compared to \emph{mode}, indicates that the former is more sensible to temporal correlations in the data, which may lead to erroneous classifications.

\subsection{Impact of Ambiguous Feature Vectors}

Here we give insights on the impact of ambiguous feature vectors in the dataset on the classification process by excluding them from the training and test sets. Not all results are shown for space reasons. With the RF classifier, overall F-scores are 0.99 both with and without shuffling, leading to a slightly better classification performance compared to the entire dataset. In the Antarex dataset, around 20\% of the feature vectors are ambiguous. With respect to this relatively large proportion, the performance gap described above is small, which proves the robustness of our detection method. In general, the proportion of ambiguous feature vectors in a dataset depends primarily on the length of the aggregation window, and on the frequency of state changes in the HPC system. More feature vectors will be ambiguous as the length of the aggregation window increases, leading to more pronounced adverse effects on the classification accuracy. 

\begin{figure}[t!]
 \centering
 \captionsetup[subfigure]{}
    \subfloat[ddot.]{
    \includegraphics[width=0.46\textwidth,trim={0 0 0 0}, clip=true]{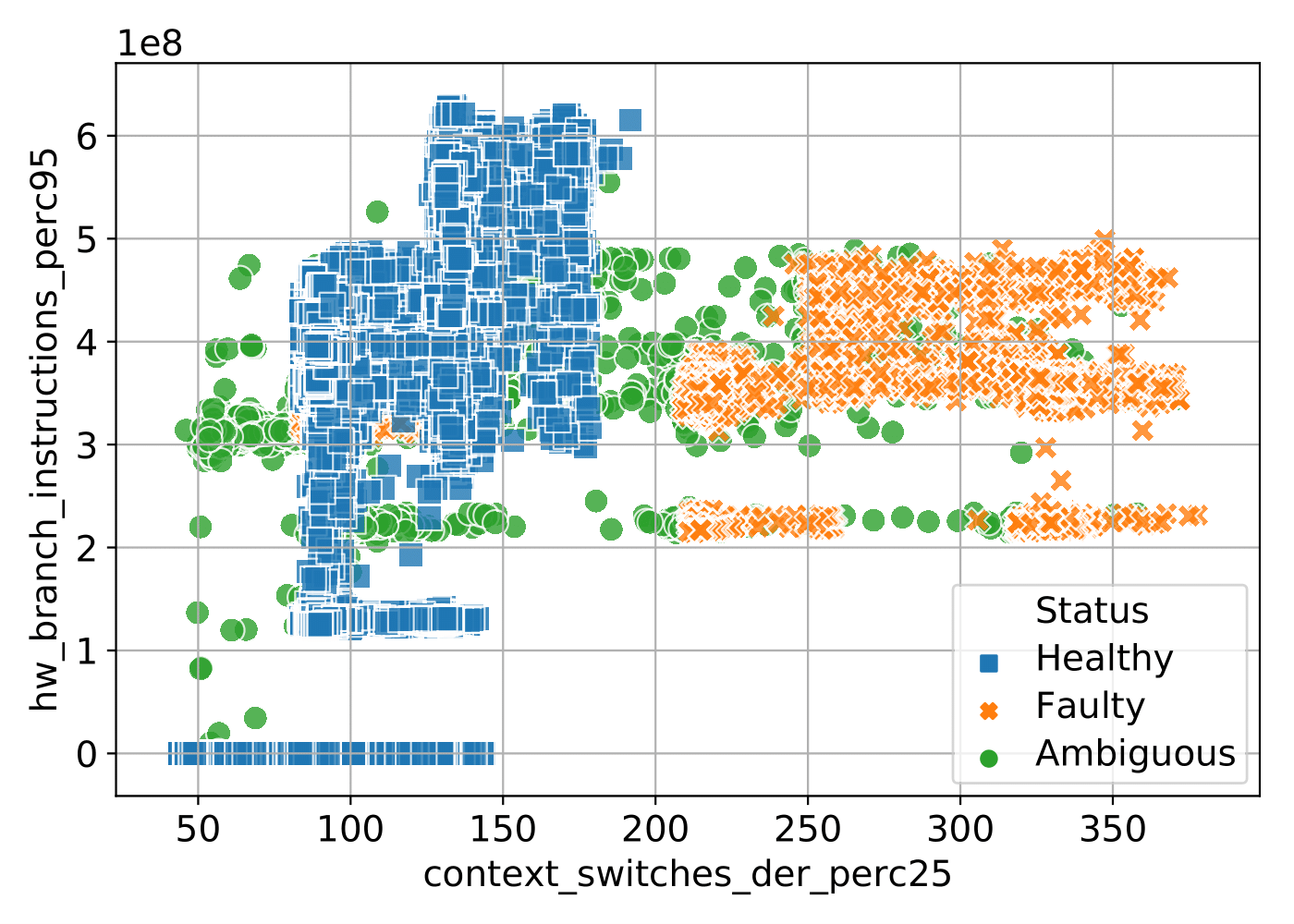}
  	}
  \subfloat[cpufreq.]{
    \includegraphics[width=0.46\textwidth,trim={0 0 0 0}, clip=true]{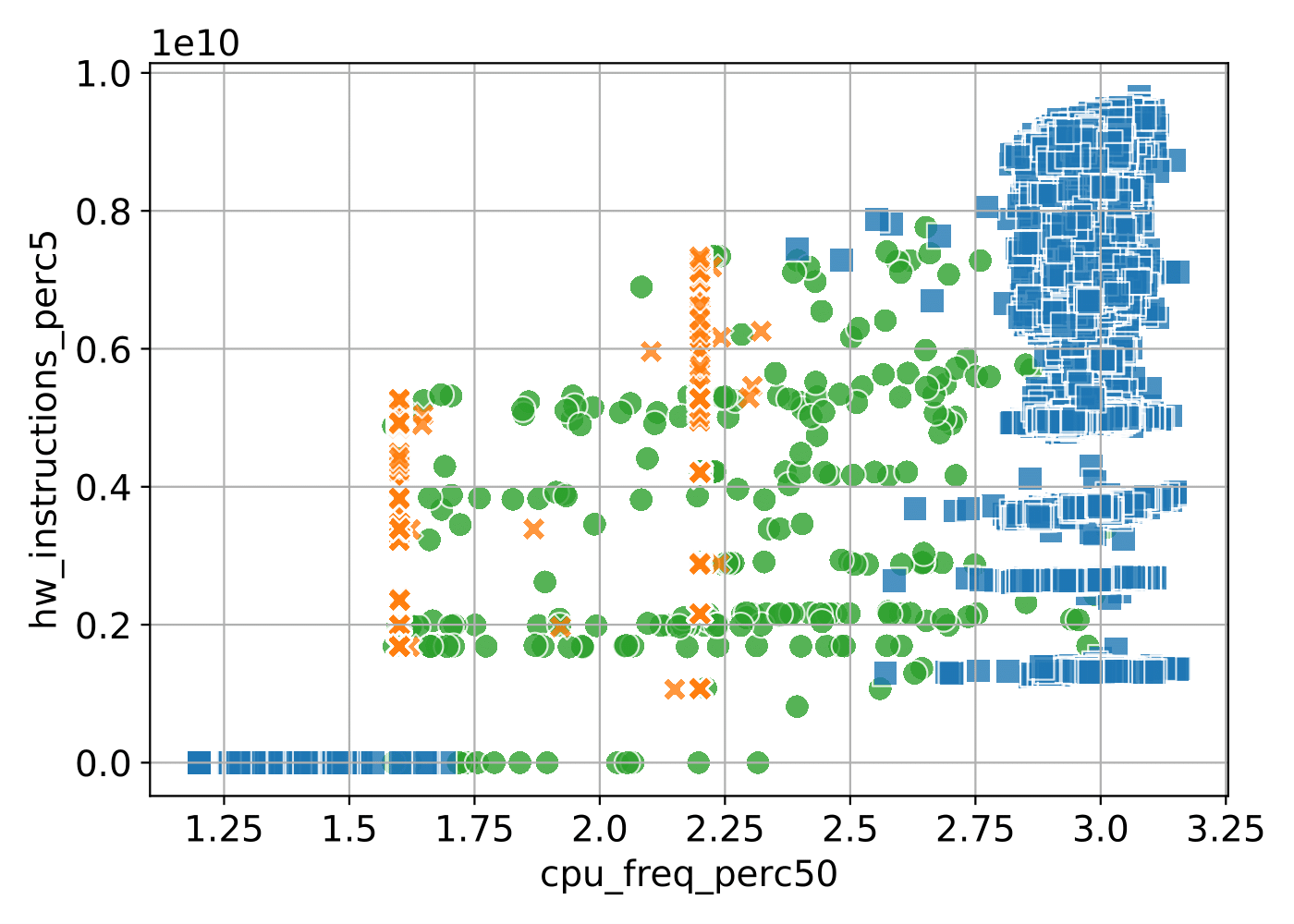}
  }
  \\
  \subfloat[memeater.]{
    \includegraphics[width=0.46\textwidth,trim={0 0 0 0}, clip=true]{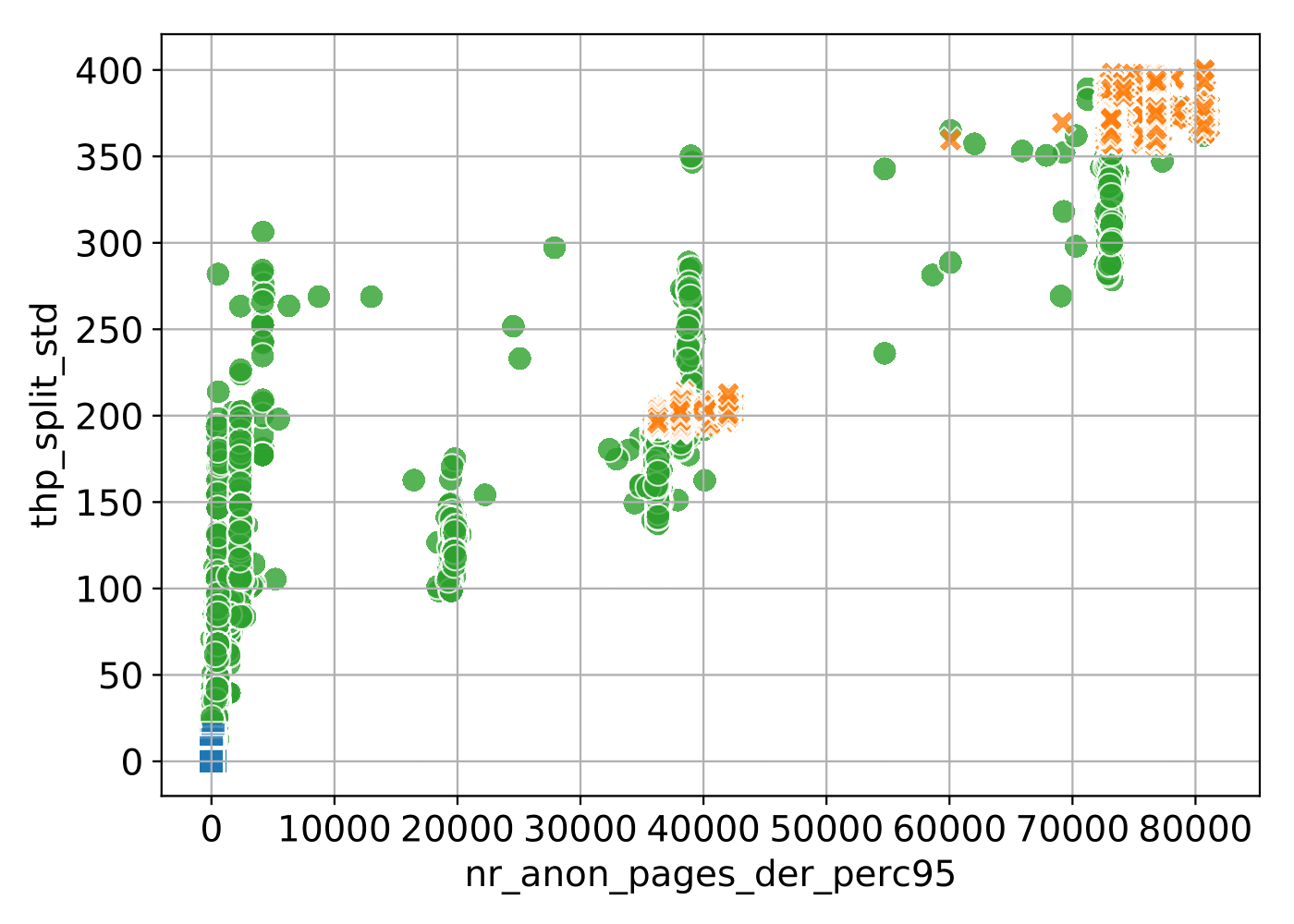}
  }
  \caption{The scatter plots of two important metrics, as quantified by a DT classifier, for three fault types. The ``healthy'' points are marked in blue, while fault-affected points in orange, and the points related to ambiguous feature vectors in green.}
  \label{results:scatter_plots}
 \end{figure}

A more concrete example of the behavior of ambiguous feature vectors can be seen in Figure~\ref{results:scatter_plots}, where we show the scatter plots of two important metrics (as quantified by a DT classifier) for the feature vectors related to the ddot, cpufreq and memeater fault programs, respectively. The ``healthy'' points, marked in blue, and the fault-affected points, marked in orange, are distinctly clustered in all cases. On the other hand, the points representing the ambiguous feature vectors, marked in green, are sparse, and often fall right between the ``healthy'' and faulty clusters. This is particularly evident with the cpufreq fault program in Figure~\ref{results:scatter_plots}b.

\subsection{Remarks on Overhead}

Quantifying the overhead of our fault detection framework is fundamental to prove its feasibility on a real online HPC system. LDMS is proven to have a low overhead at high sampling rates~\cite{agelastos2014lightweight}. We also assume that the generation of feature vectors and the classification are performed locally in each node, and that only the resulting fault diagnoses are sent externally. This implies that the hundreds of performance metrics we use do not need to be sampled and streamed at a fine granularity. We calculated that generating a set of feature vectors, one for each core in our test node, at a given time point for an aggregation window of 60 seconds takes on average 340 ms by using a single thread, which includes the I/O overhead of reading and parsing LDMS CSV files, and writing the output feature vectors. Performing classification for one feature vector using a RF classifier takes on average 2 ms. This results in a total overhead of 342 ms for generating and classifying feature vectors for each 60-seconds aggregation window, using a single thread, which is acceptable for online use. Such overhead is expected to be much lower in a real system, with direct in-memory access to streamed data, since no CSV files must be processed and therefore no file system I/O is required. Moreover, as the single statistical features are independent from each other, data processing can be parallelized on multiple threads to further reduce latency and ensure load balancing across CPU cores, which is critical to prevent slowdown for certain applications.

%% file: sections/Conclusions.tex
\section{Conclusions}
\label{section:conclusions}

We have presented a fault detection and classification method based on machine learning techniques, targeted at HPC systems. Our method is designed for streamed, online data obtained from a monitoring framework, which is then processed and fed to classifiers. Due to the scarcity of public datasets containing detailed information about faults in HPC systems, we acquired the Antarex dataset and evaluated our method based on it. Results of our study show almost perfect classification accuracy for all injected fault types, with negligible computational overhead for HPC nodes. Moreover, our study reproduces the operating conditions that could be found in a real online system, in particular those related to ambiguous system states and data imbalance in the training and test sets.

As future work, we plan to deploy our fault detection framework in a large-scale real HPC system. This will involve the development of tools to aid online training of machine learning models, as well as the integration in a monitoring framework such as Examon~\cite{beneventi2017continuous}. We also need to better understand our system's behavior in an online scenario. Specifically, since training is performed before HPC nodes move into production (i.e., in a test environment) we need to characterize how often re-training is needed, and devise a procedure to perform this.

\paragraph{Acknowledgements.} A. Netti has been supported by the \textit{Oprecomp-Open Transprecision Computing} project. A. S\^irbu has been partially funded by the EU project \textit{SoBigData Research Infrastructure --- Big Data and Social Mining Ecosystem} (grant agreement 654024). We thank the Integrated Systems Laboratory of ETH Zurich for granting us control of their Antarex HPC node during this study.